\documentclass[showpacs,twocolumn]{revtex4}
\emergencystretch=\maxdimen
\usepackage{graphicx}
\usepackage{dcolumn}
\usepackage{amsmath}
\usepackage[latin1]{inputenc}
\usepackage{graphicx, psfrag}
\usepackage{amssymb}
\usepackage[colorlinks=true, citecolor=blue, urlcolor = blue, linkcolor= red, bookmarks=true]{hyperref}
\usepackage{float}
\usepackage{amsfonts}
\usepackage{hyperref}
\usepackage{subfigure}
\usepackage{pgfplots}
\usepackage{epstopdf}
\usepackage{booktabs}

\makeatletter
\def\btt#1{\texttt{\@backslashchar#1}}
\DeclareRobustCommand\bblash{\btt{\@backslashchar}} \makeatother

\begin{document}

\title[]{Dynamical Wormholes in Higher Dimensions and the Emergent Universe}
\author{B. C. Paul}\email{bcpaul@associates.iucaa.in}

\affiliation{ Department of Physics, University of North Bengal, Siliguri, Dist. : Darjeeling 734 014, West Bengal, India \\
            and \\
            IUCAA Centre for Astronomy Research and Development, North Bengal  }

\date{\today}

\begin{abstract}
We present dynamical wormholes  in higher dimensions which admit flat emergent universe  (EU) model in an elegant way. The EU model was proposed to get rid of some of the problems of the  Big Bang cosmology. It  is free from initial singularity with other observed features of the universe.  The basic assumption of EU model was that the present universe emerged out from a static Einstein universe.  In the paper we study EU model in four and in higher dimensions and proposed that the EU originates from a   dynamical wormhole and  the throat of the wormhole is the seed of the Einstein Static universe. The shape function  obtained here admits closed, open and flat universe in  higher dimensions.  A class of new  cosmological solutions in a higher dimensional flat universe is obtained. We obtain shape functions for flat, asymptotic closed and open universe.The shape function obtained here for asymptotic closed universe is new. It is found that flat EU with a non-linear equation of state (EoS) can be accommodated with dynamical wormhole.  The non-linear EoS corresponds to   three types of  fluids in the universe.  The EoS parameter is playing an important role in determining the cosmic fluids. The space-time dimensions determines the rate of change of  a particular fluid that varies with the scale factor of a dynamically evolving  universe with non-interacting fluids. Considering  interaction  at time $t > t_0$,  among the three types of fluids it is possible to describe the observed universe satisfactorily.  In a higher dimensional universe it is found that near the throat null energy condition (NEC) is violated, but away from the throat NEC is found to obey admitting the observed universe for a flat case. Another interesting aspect of the EU model is that it permits late accelerating phase. However, in asymptotic closed or open universe, flat emergent universe can be accommodated with NEC which is obeyed right from the throat to the present epoch. The tension at the throat of the  wormhole is estimated  which is found to depend on the initial size of the Einstein static universe and dimensions of the universe. It is interesting to note that NEC is not violated  to accommodate dynamical wormholes for closed or open universe. Although exotic matter  is required at the throat for the flat universe,  no exotic matter is  required for  closed or open universe which encompass the  emergent universe.

\end{abstract}

\maketitle

\section{Introduction}
\label{intro}
In recent years,  there has been a growing interest to study non-trivial spacetime topologies in Einstein's General Theory of Relativity (GTR) in understanding different properties of the universe. Lorentzian wormhole solutions are  permitted in GTR in various conditions. The  wormholes  allow topological passage through hypothetical tunnel like bridges connecting two distant regions of a universe or  two different universes.    In 1957  Wheeler and Misner \cite{1,2} coined the term  {\it wormhole}. A number of interesting features of the traversable Lorentzian wormholes led to a spurt in activities in theoretical studies \cite{3,4}. 
The Lorentzian wormholes are handles in the spacetime topologies linking widely separated regions of the universe.It was demonstrated that weak energy condition is violated at the throat of the wormholes. Visser \cite{5a,5b,5c,5d} elegantly constructed traversable wormhole based on surgical modified solution of the Einstein's field equations. 
 The observed features of the Lorentzian wormhole solution is acceptable for constructing  a hypothetical time machines \cite{5,6}.  In  the literature \cite{7,8,9},  it  is  found that astrophysical accretion of ordinary matter could convert wormholes into black holes. Hayward \cite{10} shown that black holes and wormholes are interconvertible structures and stationary wormholes may be the final stage of a evaporating black hole. 
The stability of the wormhole solutions are analysed in detail in the presence of different types of matter. The wormholes may have played an important role in the early phase of the universe. 
Considering the end-state of stellar-mass binary black hole mergers in GW150914, a  two parameter family of wormholes in  a scalar tensor theory,  it is shown \cite{11}    that one may detect the signatures  present in the emitted gravitational waves as they settle down in the post-merger phase from an initially perturbed state.

 It is known that wormhole solution in GTR exists when the energy conditions of the matter is violated. The energy-momentum tensor of the matter sector of the Einstein gravity supporting such geometries violates the null energy condition (NEC) at least near the throat region of the wormhole \cite{12,13,14} for its existence.  Such violation of energy condition indicates the presence of exotic matter.  
 The recent  predictions of the astronomical observations  is very interesting in the subject  as it  stirring an elegant discovery that the universe is experiencing a phase of accelerated expansion.  This is a great challenge in theoretical physics to explore the fields/matter which can not be understood in the framework of standard model of particle physics.  It is known that in describing both the late accelerating  phase of the universe and the wormholes one requires a situation allowing  the null energy conditions (NECs) to violate. Thus there is an amazing and unexpected overlapping that occurs between these two different topics. The matter sector in this case is considered to be composed of matter different from normal matter. In cosmology to understand the late acceleration of the universe matter sector is modified with dark energy, phantom, Chaplygin gas \cite{15,16,17,17a,17b}.  In the literature,  the  phantom field which is an exotic matter  considered to realize wormhole geometry \cite {18,19,20}.   The matter with  property  - $p_{eff}|_{r(0)}> \rho_{eff}|_{r(0)}>0$ (where $r(0)$ is the throat radius of the wormhole) is called exotic matter.  The  exotic matter at the throat of the wormhole signifies that an observer who moves through the throat with a radial velocity approaching  the speed of light will observe presence of negative energy density leading to the violation of the null, weak, strong energy conditions. However, it will be interesting if one can  construct an exact traversable wormhole without exotic matter. 
 
 Euclidean wormholes are important  in understanding origin of the universe without singularity  which contributes to the Euclidean path integral formulation of quantum gravity and quantum cosmology.  A consistent theory of quantum gravity is  not yet known.  
 However, the superstring theory is considered to be a promising candidate for quantum gravity
which may unify gravity with the other fundamental forces in nature which requires 10
dimensions for its consistent formulation. The idea that spacetime dimensions should be extended from four to higher dimensions originated from the work of Kaluza  \cite{21} and Klein \cite{21a}  (KK) who first tried to unify gravity with electromagnetism. But the old KK-approach does not work well. The higher dimensional theories revived once again in recent decades due to the advent of string theory which led a paradigm shift in
higher-dimensional cosmology. It is therefore important to generalize the known results in four dimensions in the framework of Einstein's GTR in higher dimensions. Lorentzian wormhole solution obtained in a four dimensions with $R=0$ is studied in  four dimensions \cite{22}. It is shown \cite{23}-\cite{23e} that the NEC, or more precisely the
averaged null energy condition can be avoided in certain regions of dynamical wormholes. 
In this case wormholes in a time-dependent inflationary background \cite{24,24a} are employed   to enlarge an initially small and possibly submicroscopic wormhole encompassing an early inflationary universe scenario. Finally, it  will continue to be enlarged by the subsequent Friedmann-Robertson-Walker (FRW) phase of expansion. One could
perform a similar analysis as used in Ref. (\cite{24}), replacing the de Sitter scale factor by a FRW scale factor (\cite{23b}-\cite{23d}). In particular, in Ref. (\cite{23b}-\cite{23c}) specific examples for evolving wormholes that
exist only for a finite time are considered, and analyzed for a special class of scale factors that permits violation of the weak energy condition (WEC)  for wormholes. 
Higher-dimensional evolving wormholes satisfying the null energy condition was studied by Zangeneh {\it et al.} \cite{22a},  considering a particular class of wormhole solutions corresponding to the choice of a spatially homogeneous
Ricci scalar. They explored wormholes with normal and exotic matter and obtained a number of 
wormhole solutions including those in four dimensions that satisfy the null energy condition. In five dimensions it is found that the solutions  satisfy the null energy condition throughout either power law or de Sitter evolution.

Brane world \cite{eu5,eu6}, Brans-Dicke theory \cite{eu7} and in the context of a non-linear
sigma model \cite{eu8}. EU scenario accommodates a  late time de Sitter  expansion which 
naturally permits late time acceleration of the universe. The EU scenario is promising from the
perspective of offering  unification of early as well as late time dynamics of the universe. It may be noted
that unification in such emergent universe model lies in the choice of equation of state for the polytropic fluid. A number of issues pertaining to different models of EU have been discussed in the literature \cite{eu9,eu10}.
The EU model with a polytropic non-linear equation of state (EoS) \cite{eu4} gives rise to a
flat universe with a composition of three different types of cosmic  fluids determined by the EoS parameters
$A$ and $B$. In the model one problem was that the fluid contents of the universe is of definite type once EoS parameter $A$ is fixed. The three fluids are identified with exotic matter, dark energy and a barotropic fluid. 
Later using interactive fluid models it is shown \cite{eu11,eu12} that a viable EU scenario
can be realized. The basic need of an emergent universe is the existence of a  Einstein's static Universe phase in the infinitely past time which however emerge out to an expanding universe. But it is not known how a static Einstein universe exists in the infinitely past time. Therefore, it is of interest to explore the origin of an initial static Einstein phase of the universe which is essential to realize  EU scenario. The wormhole physics will be employed here. The present paper investigates the possibility and naturalness of expanding wormholes in higher dimensions,
which is an important ingredient of the modern theories of fundamental physics, such as string theory, supergravity,
Kaluza-Klein, and others.  The higher-dimensional spacetimes is an
important ingredient of modern theories of fundamental
physics. In this context, the existence of higher dimensions may help to construct wormhole solutions that permits EU model.  The motivations for the paper is to obtain dynamical wormhole that permits an expanding cosmological model which accommodates emergent universe scenario \cite{eu4}-\cite{eu10} in higher dimensions.


The paper is organised as follows: In sec.II, the Einstein field equation in higher dimensions is given.
 In sec.III, we have considered the static spherically symmetric metric to describe the wormhole geometry and the necessary conditions which are to be satisfied  making use of  shape function obtained from a homogeneous Ricci scalar ($R$).  The field equations are determined for the wormhole metric. In sec. IV, we have considered two possible wormhole shape functions with their features, and in sec. V, the physical analysis has been carried out. The validity of NEC and WEC are studied by plotting graph. In sec. VI, the constraints on the tension and mass density at the throat in Higher dimensions is given. The results are summarised in sec. VI| followed by a brief discussion.


\section{The Gravitational Action and Solutions in Higher Dimensions}

The  gravitational action  in D-dimensions is given by 
\begin{equation}
\label{e1}
I=  \int d^{D}x\sqrt{-g} \left( \frac{1}{2} R + {\it L_m} \right)
\end{equation}
where $R$ is the  Ricci scalar curvature and $ {\it L_m} $ is the matter Lagrangian, we consider $ c=8 \pi G_D=1$. Varying the action with respect to the metric, we obtain $D$-dimensional Einstein field equation
\begin{equation}
\label{e2}
R_{AB} -\frac{1}{2} g_{AB}R = T_{AB}
\end{equation}
where $A, B = 0,1, ... D-1$ and  $T_{AB }$ is the matter stress-energy tensor. For an expanding wormhole solutions we use the metric
\begin{equation}
\label{e3}
ds^{2}= - dt^{2}-  S(t) ^2 \left[\frac{dr^{2}}{1- \alpha(r) } + r^2 d\Omega^2_{D-2} \right]
\end{equation}
where $ d\Omega^2_{D-2}=  d\theta_1^2+ sin^2 \theta_1 d\theta_2^2 +sin^2 \theta_2 ( d\theta_3^2 + ... +sin^2 \theta_{D-3} \; d\theta_{D-2}^2 $,$S(t)$ represents scale factor, $\alpha(r)$ is an unknown dimensionless function, defined as $\alpha(r)= \frac{b(r)}{r}$ where $b(r)$ denotes the shape function \cite{4,5}. The metric is a generalization of Friedmann-Robertson metric when $\alpha(r) =0$.  It may be noted that when the dimensionless shape function $\alpha(r) \rightarrow 0$, the metric reduces to a flat Friedmann-Robertson metric. It approaches to the static wormhole metric as $S(t) \rightarrow constant$. In this case the wormhole form of the metric is preserved
with time. Let us consider an embedding of $t=const.$ and  $\theta_{D-2}=\frac{\pi}{2}$ slices of the space given by eq. (\ref{e3}), in a flat three dimensional Euclidean space with metric becomes
\begin{equation}
\label{e4}
ds^2=d\bar{z}^2+d\bar{r}^2+\bar{r}^2 d\phi^2.
\end{equation}

The metric of the wormhole slice is given by
\begin{equation}
\label{e5}
ds^{2}= S(t) ^2 \left[\frac{dr^{2}}{1- \alpha(r) } + r^2 d\phi^2 \right].
\end{equation}
Comparing eq. (\ref{e4}) with eq. (\ref{e5}) one can write
\[
\bar{r} = S(t) r|_{t=const.}  ,
\]
\begin{equation}
\label{e6}
d\bar{r}^2 = S^2(t) dr^2|_{t=const.},
\end{equation} 
they may be regarded as  rescaling of $r$-coordinate on each $t=const.$ slice \cite{25}.

Now one obtains a metric using in eq. (\ref{e4}) which is given by
\begin{equation}
\label{e6a}
ds^{2}=\frac{d\bar{r}^{2}}{1- \bar{\alpha}(\bar{r}) }  + \bar{r}^2 d\phi^2,
\end{equation}
where $\bar{\alpha} (\bar{r}_{0})=1$, $i.e.$ $\bar{b} (\bar{r})$ attains a minimum at $\bar{b} (\bar{r}_{0})=\bar{r}_{0}$.
Using eq. (\ref{e6}) and eq. (\ref{e7}) one gets
\begin{equation}
\label{e7b}
\bar{\alpha} (\bar{r}) = S(t) \alpha (r).
\end{equation} 
In the case of a evolving wormhole having same size and shape relative to the ($\bar{z} , \bar{r}, \phi$ ) coordinate system will have that in the initial ($z,r,\phi$) embedding space coordinate system. Using eqs. (\ref{e4}) and (\ref{e6a})  one obtains 
\begin{equation}
\label{e7a}
\frac{d\bar{z}}{d\bar{r}}=\frac{dz}{dr}
\end{equation}
with $\bar{z}(\bar{r}) = \pm \; S(t) \; z(r)$. Now the connection of the embedding space at any time $t$ and the initial embedding space at $t=0$ is  given by
\[
\hspace{1.5 cm}  ds^{2}= d\bar{z}^2 + d\bar{r}^2 +   \bar{r}^2 d\phi^2 
\]
\begin{equation}
\label{e8}
\hspace{2.0 cm} = R^2(t) \left( dz^2 +dr^2 +r^2 d\phi^2 \right).
\end{equation}
The wormhole will change relative to the initial $t=0$ embedding space. The flaring out condition for the evolving wormhole at or near the throat is
\begin{equation}
\label{e8a}
\frac{d^2 \bar{r} (\bar{z})}{d\bar{z}^2} > 0.
\end{equation}
Using eqs. (\ref{e6}), (\ref{e7b}) and  (\ref{e7a}) we get
\begin{equation}
\label{e9}
\frac{d^2 \bar{r} (\bar{z})}{d\bar{z}^2} = \frac{1}{S(t)} \left( - \frac{\alpha'}{2 \alpha^2} \right) = \frac{1}{S(t)} \frac{d^2r(z)}{dz^2} > 0.
\end{equation}
where prime represents derivative w.r.t. $r$. It is evident that flaring out condition has the same form in both the embedding spacetimes considered here. Thus a traversable wormhole can be represented by the metric given by 
eq. (\ref{e2}) provided it satisfies the following conditions \cite{4} :
\begin{equation}
\label{e10}
\alpha (r_0)=1, \; \; \; \; \alpha (r) < 1, \; \; \; \; \alpha'(r) < 0
\end{equation}
where $r_0$ represents the wormhole throat.

Using the metric (2) and the energy momentum tensor $T^{A}_{B} = diag \left( - \rho, P_r, P_t, P_t, ...... \right)$,  Einstein field equations become
\[
\rho(r, t)= \frac{ (D-2)(D-3) \; \alpha(r)}{2 S^2(t) r^2} + \frac{(D-2) \; \alpha'(r)}{2 S^2(t) r}
\]
\begin{equation}
\label{e11}
\hspace{1.5 cm}  + \frac{(D-1)(D-2)}{2} \frac{ \dot{S}^2(t)}{S^2(t)},
\end{equation}
\[
P_r (r, t)= -  \;(D-2)  \frac{ \ddot{S}(t)}{S(t)} - \frac{(D-2)(D-3)}{2}   \frac{\dot{S}^2(t)}{S^2(t) }
 \]
\begin{equation}
\label{e12}
\hspace{1.5 cm}  - \; \frac{(D-2)(D-3) \alpha(r)}{2 r^2 S^2(t)},
\end{equation}
\[
P_t (r, t)=   -\; (D-2) \frac{\ddot{S}(t)}{S(t)}  -  \frac{(D-2)(D-3)}{2}   \frac{\dot{S}^2(t)}{S^2(t) }
 \]
\begin{equation}
\label{e13}
\hspace{1.5 cm}  - \; \frac{(D-3)(D-4) \alpha(r)}{2 r^2 S^2(t)} -  \;  \frac{(D-3) \alpha'(r)}{2 r S^2(t)}
\end{equation}
where over dot represents derivative with respect to time.
The Ricci scalar is given by
\[
R= 2 (D-1)  \frac{ \ddot{S}(t)}{S(t)} + \frac{(D-1)(D-2)}{2}   \frac{\dot{S}^2(t)}{S^2(t) }
 \]
\begin{equation}
\label{e14}
\hspace{1.5 cm}  + \frac{(D-2)(D-3) \alpha(r)}{r^2 S^2(t)} + \frac{(D-2) \alpha'(r)}{r S^2(t)}.
\end{equation}
In this paper we look for relativistic  wormhole solutions corresponding to the homogeneous Ricci Scalar, $i.e.$ $\frac{\partial R}{\partial r} =0$ which leads to a second order differential equation in $\alpha$ given by
\begin{equation}
\label{e15}
r^2    \alpha''(r) + (D-4) r \alpha'(r) -2 (D-3) \alpha (r) =0.
\end{equation}
The above second order differential equation has solution
\begin{equation}
\label{ev}
 \hspace{1.8 cm}\alpha(r) = C_1 \; r^{3-D} \pm C_2 \; r^2
\end{equation}
where $C_1$ and $C_2$ are integration constant. Using the constraint $\alpha (r_0)=1$,
one obtains the following solutions given by
\begin{equation}
\label{e16}
Case \; (I) \hspace{1.5 cm} \alpha (r) = \frac{ r_0^{D-3}   - k \; r_0^{D-1}}{r^{D-3}} + k\; r^2
\end{equation}
 for plus sign in eq. (\ref{ev}) and 
\begin{equation}
\label{e17}
Case \; (II) \hspace{1.5 cm} \alpha (r) = \frac{ r_0^{D-3}   + k \; r_0^{D-1}}{r^{D-3}} - k\; r^2
\end{equation}
for minus sign in eq. (\ref{ev}) where we redefined  $C_2=k$ and $C_1=\frac{1- k r_0^2}{ r_0^{3-D}} $ so as to make it convenient to identify the curvature index $k$ of the Robertson-Walker metric.
 It is evident that although $k$ is a continuous variable, it leads to the spacetime which is asymptotically FRW and applied the normalization that $k= 0, -1$  for Case (I) and $k=1$ for Case (II). In the absence of a cosmological constant the Ricci Scalar may be zero in GTR to obtain a static wormhole \cite{22,23c}. On the other hand a vanishing  Ricci scalar is obtained if the scale factor is independent of time accommodating the  solutions given in Case (I) and Case (II).  The dimensionless shape function $\alpha (r)$ should satisfy the condition mentioned in eq. (\ref{e10}), it is found that the solutions obtained in Case (I) represents flat or open universe for $k=0, -1$ respectively but not the closed universe but in the Case (II) a closed universe ($k=1$) is permitted. Thus a new class of solutions in the later case is  obtained for a closed universe accommodating wormholes. In this case two closed universes are connected by a wormhole.
 
A flat universe can be represented by $k=0$, the field 
eq. (\ref{e11})- eq. (\ref{e13}) can be rewritten as
\begin{equation}
\label{e18}
\rho =\rho_{fb}
\end{equation}
\begin{equation}
\label{e19}
P_r = - \; \frac{(D-2)(D-3) \; r_0^{D-3}}{2r^{D-2} S^2} + P_{fb}
\end{equation}
\begin{equation}
\label{e20}
P_t =  \frac{(D-3) r_0^{D-3}}{2r^{D-2} S^2} + P_{fb}
\end{equation}
where we denote  the flat ($k=0$) background components $\rho_{fb}$ and $P_{fb}$ as
\begin{equation}
\label{e21}
\rho_{fb}= \frac{(D-1)(D-2)}{2} \frac{\dot{S}^2}{S^2}
\end{equation}
\begin{equation}
\label{e22}
P_{fb}=  - \;  (D-2) \frac{\ddot{S}}{S} - \frac{(D-2)(D-3)}{2} \frac{\dot{S}^2}{S^2}
\end{equation}
respectively. It corresponds to a universe with isotropic cosmic fluid.

In the open universe background ($k=-1$), we get
\begin{equation}
\label{e23}
\rho =\rho_{ob},
\end{equation}
\begin{equation}
\label{e24}
P_r = - \; \frac{(D-2)(D-3) (r_0^{D-1} + r_0^{D-3})}{2r^{D-1} S^2} + P_{ob}
\end{equation}
\begin{equation}
\label{e25}
P_t =  \frac{(D-3) (r_0^{D-1} + r_0^{D-3})}{2r^{D-1} S^2} + P_{ob}
\end{equation}
where $\rho_{ob}$ and $P_{ob}$ corresponds to the open background which are
\begin{equation}
\label{e26}
\rho_{ob} = \rho_{fb} - \frac{(D-1)(D-2)}{2S^2},
\end{equation}
\begin{equation}
\label{e27}
P_{ob}=P_{fb} + \frac{(D-2)(D-3)}{2S^2}.
\end{equation}

A closed universe background $k=1$ solutions obtained here  corresponding to minus sign in eq. (\ref{ev}) is a new set of solutions with wormholes.  Thus in this case we can represent following:
\begin{equation}
\label{e28}
\rho =\rho_{cb},
\end{equation}
\begin{equation}
\label{e29}
P_r = - \; \frac{(D-2)(D-3) (r_0^{D-1} + r_0^{D-3})}{2r^{D-1} S^2} + P_{cb}
\end{equation}
\begin{equation}
\label{e30}
P_t =  \frac{(D-3) (r_0^{D-1} + r_0^{D-3})}{2r^{D-1} S^2} + P_{cb}
\end{equation}
where $\rho_{cb}$ and $P_{cb}$ corresponds to the closed universe background which are
\begin{equation}
\label{e31}
\rho_{cb} = \rho_{fb} - \frac{(D-1)(D-2)}{2S^2},
\end{equation}
\begin{equation}
\label{e32}
P_{cb}=P_{fb} + \frac{(D-2)(D-3)}{2S^2}.
\end{equation}
It is evident that wormholes are obtained with anisotropic fluid. In the case of flat universe at a large distance away from the throat it admits isotropic pressure. We get similar picture if the universe is non-flat asymptotically.

\section{Wormholes Solutions }

In this section we consider the parameter: $\alpha (r) = \frac{b(r)}{r}$, where $b(r)$ is  typical shape function which satisfy the following constraints for accommodating wormholes: \\

$\bullet$ The range of  radial coordinate $r$  is $r_{0} \le r \le \infty$, with $r_{0}$ being the throat radius.\\

$\bullet$ The shape function $(b(r))$ satisfies the condition $b(r_{0})=r_{0}$, at the throat and away from the throat $\it{i.e.}$ for $r>r_{0}$ it must satisfy the constraint condition
\begin{equation}
\hspace{2.5 cm} 1-\frac{b(r)}{r} > 0.
\end{equation}

$\bullet$ For a physical  flaring out condition, it must be satisfied by the shape function $b(r)$ which at the throat 
of a wormhole solution becomes $\it{i.e.}$ $b'(r_{0})<1$.

$\bullet$ For an asymptotic flatness of the spacetime geometry one obtains
\begin{equation}
\label{e33}
\; \hspace{2.0 cm} \frac{b(r)}{r} \rightarrow 0 \ \ \ as \ \ \ |r| \rightarrow \infty
\end{equation}

 It is known that the matter that supports the static wormhole geometry in Einstein's GTR violates the Null Energy condition (NEC):
 $T_{AB} U^{A} U^{B} \geq 0$ where $U^{A}$ is a null vector. \\
 
 1. NEC can be expressed in terms of energy density and pressure as
\begin{equation}
\label{e34}
\hspace{1.8 cm} \rho + P_r \geq 0,  \; \; \; \; \; \; \rho+ P_t \geq 0. 
\end{equation}
The other energy conditions \cite{26}   are \\

2. DEC: $\rho \ge   | P_i |$, \\                  

3. WEC : $\rho \ge 0$; \;  $\rho +P_{i}>0$;\\

4. SEC: $\rho+P_{r}+2P_{t}\ge 0$;\\

where $i=r,t.$\\

{\bf A. Flat Background}

We consider a non-linear equation of state for matter as follows :
\begin{equation}
\label{e35}
\hspace{1.8 cm}  P_r = A \rho -\frac{B}{\rho^{\epsilon}}
\end{equation}
where $\epsilon > 0$ corresponds to modified Chaplygin gas \cite{mcg}. In a flat background an Emergent Universe model is obtained with a non-linear equation of state \cite{eu4} which corresponds to   $\epsilon <0$. Recently  a number of theoretical aspects of EU model with  such non-linear part of the EoS is reported,  the non-linear term represents  the presence of initial viscosity in the universe \cite{non}.  In the literature \cite{others},  EU model is implemented in  a number of theories of gravity and found to work satisfactorily to encompass the observed universe. In the case of a flat background EU scenario in a flat universe  is obtained with a non-linear equation of state EoS corresponding to  $\epsilon = - \frac{1}{2}$ which is given by given by
\begin{equation}
\label{e35a}
\hspace{1.8 cm}  P_r = A \rho -B \sqrt{\rho}
\end{equation}
where $A$ and $B$ are arbitrary parameters we use them to determine the composition of matter in the universe which is interesting. Now using  the  Einstein field eqs. (\ref{e21}) and  (\ref{e22}), we get the second order differential equation given by
\begin{equation}
\label{e36}
\hspace{1.8 cm} \frac{\ddot{S}}{S} + \delta \; \frac{\dot{S}^2}{S^2} -  \beta \; \frac{\dot{S}}{S}   =0
\end{equation}
where we denote $ \delta = \frac{A \; (D-1)  -2}{2}  $ and   $ \beta= \sqrt{ \frac{(D-1)}{2(D-2)}} \; B$ for simplicity. The eq. (\ref{e36}) can be integrated once to obtain
\begin{equation}
\hspace{2.5 cm} \dot{S} \; S^{\delta} = \kappa e^{\beta  t}
\end{equation}
where $\kappa$ is an integration constant. Integrating the above equation once we obtain the solution
\begin{equation}
\label{e37}
\hspace{1.0 cm} S(t) = \left( (1+ \delta) \; \kappa_o  + \frac{\kappa \; (1+\delta)}{\beta} \; e^{\beta  t} \right)^{\frac{1}{1+ \delta}}.
\end{equation}
where $\kappa_o$ is constant of integration. We note the following :\\
case $(i)  \; $ If $B <  0$, the solution has a singularity and it is not interesting. \\

case $(ii) \; $ If $B >  0$, the solution describes an emergent universe if $\delta  >  - 1$ with positive $\kappa$ and $\kappa_o$.\\

\begin{figure}
\includegraphics[scale=0.8]{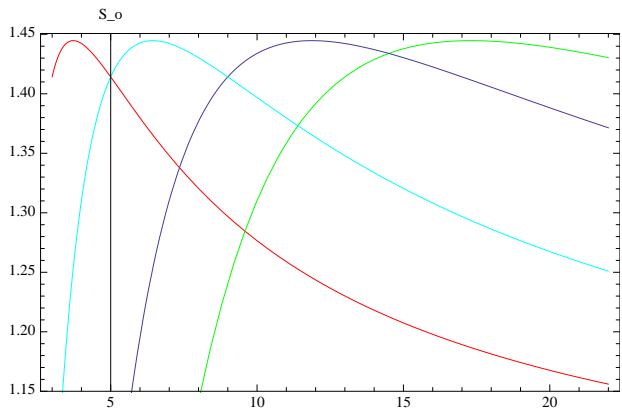}
\caption{Variation  of the initial size of the Einstein static universe with Dimension $D$ in the horizontal axis for EoS parameter $A= \frac{1}{3}\; (Green), \; \frac{1}{2}  \;  (Blue),  \; 1 \; (Cyan), \; $2$ \;(Red)$ .}
\label{Fig: 1}
\end{figure}

The solutions given in  case $(ii)$ is interesting, it leads to emergent universe scenario with asymptotic closed background. In the emergent universe there is no initial singularity and the universe evolved out from a Einstein static phase in the infinite past $i.e.$ $t \rightarrow - \infty$. It is evident that as $A$  increases the initial size $S_o$ will attain maximum at lower dimension. Thus the value of $A$ plays here an important role for determining the maximum size of the initial universe in 4-dimensions. Thus one can predict the fluids inside a Higher dimensional universe.
In the EU scenario the initial Einstein static phase is an assumption  (\cite{eu1}- \cite{eu4}).  The origin of such phase is not studied. The  wormhole physics is used to demonstrate the existence of such a phase of the universe at the throat of the wormhole. In the next section we consider case  $(ii)$   with the shape function $b(r) =   r \; \left(\frac{r_0}{r} \right)^{D-3}$ obtained from eq. (\ref{e16}). It is a positive constant $b =2  M$ (where $M$ represents mass) at $D=4$ but $b(r)$ is a decreasing function of $r$ for $D  > 4$.  Using eqs. (\ref{e18}) - (\ref{e20}) we represent the following :
\begin{equation}
\label{e38}
\rho +P_{r}=\rho_{fb} + P_{fb} - \frac{(D-2)(D-3)}{2 r^{D-1}S^2} \; r_0^{D-3},
\end{equation}
\begin{equation}
\label{e39}
\rho +P_{t}=\rho_{fb} + P_{fb} + \frac{(D-3)}{2 r^{D-1}S^2} \; r_0^{D-1},
\end{equation}
where
\begin{equation}
\label{e40}
\rho_{fb} + P_{fb} = (D-2)  \left( \frac{\dot{S}^2}{S^2} -  \frac{\ddot{S}}{S} \right).
\end{equation}
In the the case of a flat emergent universe \cite{eu4} and the solution obtained here corresponding to eq. (\ref{e40}) we get 
\begin{equation}
\label{e40a}
\hspace{1.0 cm} \rho_{fb}+P_{fb}  \neq 0.
\end{equation}
In the case of a power law and de Sitter solutions, one obtains a vanishing value \cite{25}, thus it is different in the EU model. Thus new result in emergent universe scenario leads to interesting result which we discuss in sec. 
It is evident that the Einstein static universe corresponds to $\dot{S} =0$ which leads to $S_0=\left( (1+\delta) \kappa_0\right)^{\frac{1}{1+\delta}}$ at infinitely past  ($t \rightarrow - \infty$) corresponding to the wormhole throat for $D >2$. It is found that the size of the initial Einstein static universe varies with dimensions of the universe and $A$. For a given $A$ there is a maximum $S_0$ which varies for different space-time dimensions. The maximum value however is independent of $A$. The radial variation of NEC $\rho + P_r <0$  is plotted in Fig. (2), it is violated in the past near the throat  but away from the throat NEC is obeyed.

\begin{figure}
\includegraphics[scale=0.8]{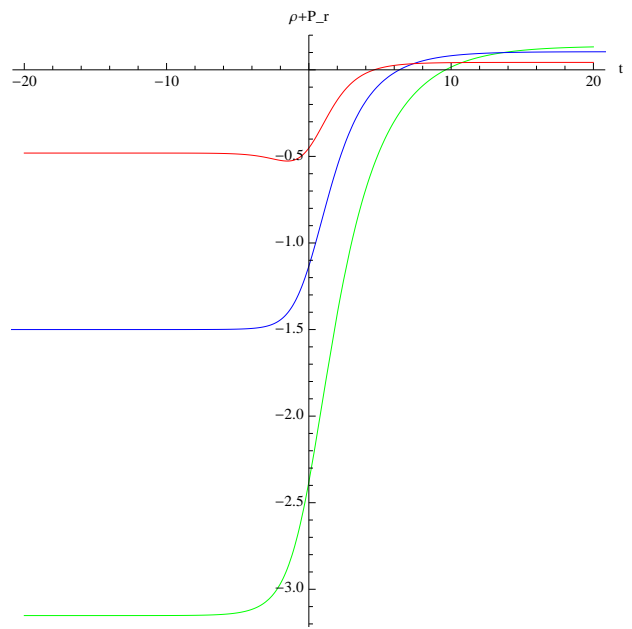}
\caption{Variation  of NEC ($\rho+p_{r})$ with time for dimension $D=4,\; 5 \,; 6$ . The red, blue and green curves correspond to $D=4,5,6$ respectively.}
\label{Fig: 1}
\end{figure}
\begin{figure}
\includegraphics[scale=0.8]{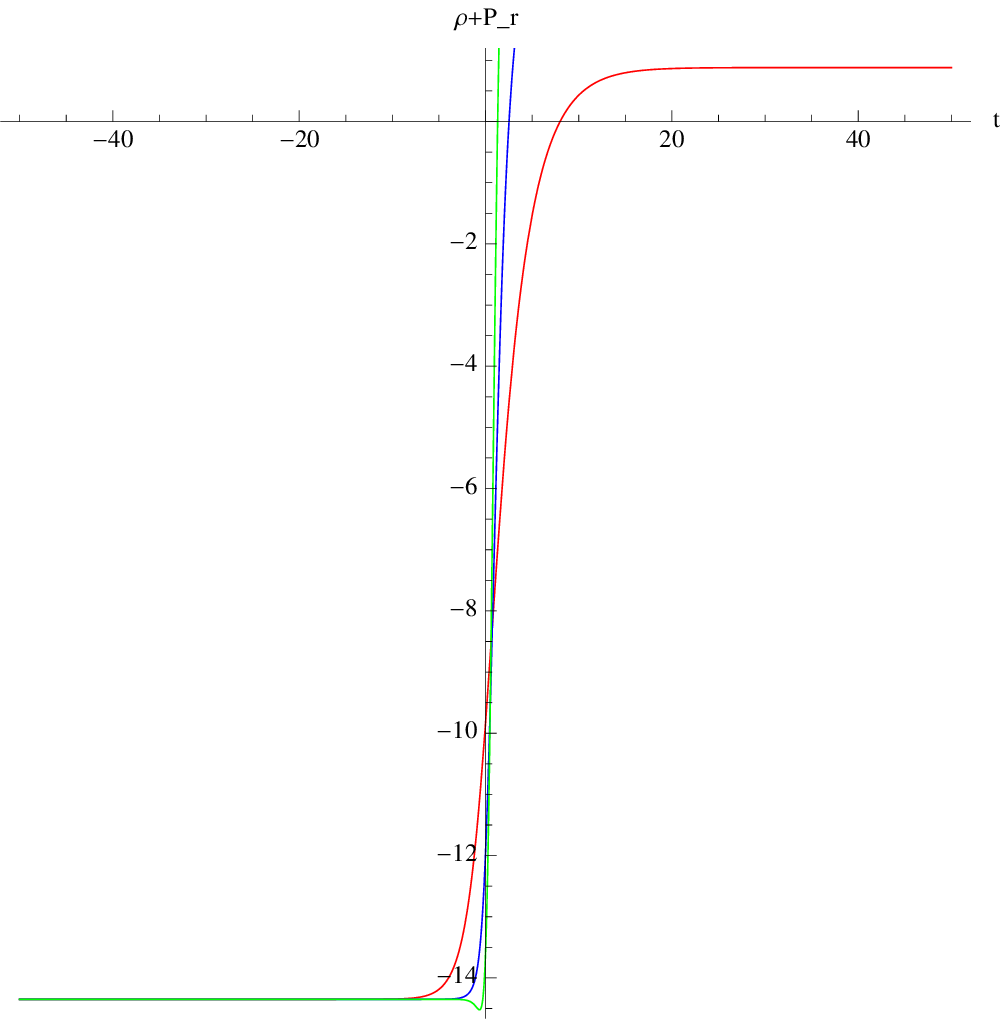}
\caption{Variation  of NEC ($\rho+p_{r})$ with time for dimension $D=10$ with $B=1,2,3$ for a given $A$. The red, blue and green curves correspond to $B=1,2,3$ respectively. }
\label{Fig: 2}
\end{figure}
\begin{figure}
\includegraphics[scale=0.8]{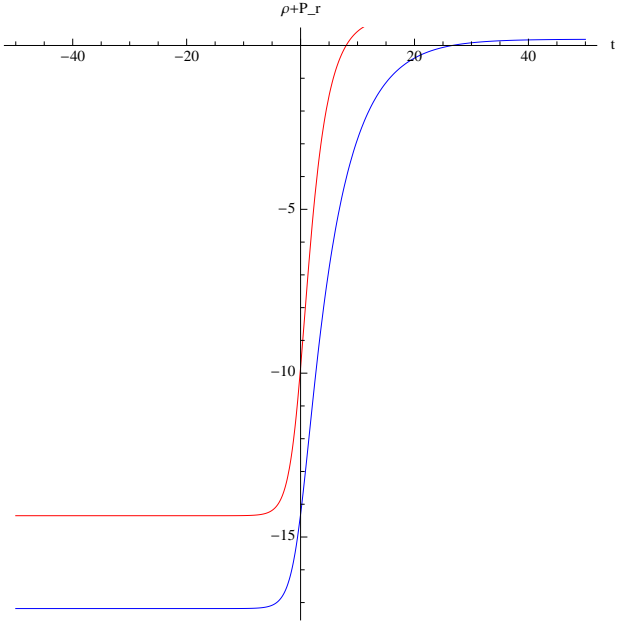}
\caption{Variation  of NEC ($\rho+p_{r})$ with time for dimension $D=10 $ with $A=1,2$ for a given $B$. The red and blue curves correspond to $A=1,2$ respectively.  }
\label{Fig: 3}
\end{figure}
It is also evident that NEC is obeyed at an early time for $D=4$ compared to other dimensions more than four with a given value of $A$ and $B$. The time evolution of NEC is plotted in Fig. 3, it is found that violation of NEC  disappears  earlier for a given $A$ when $B$ is increased in $D=10$ dimensions. However, in  Fig. 4, it is found that for a given $B$ when $A$ is increased the NEC begins to obey at an earlier epoch. A 3D plot of NEC is drawn  in Fig. 5 showing the variation with respect to $t$  and $r$ at  $D=4$ dimensions. The emergent universe at late time asymptotically attains an exponential  phase which is
\begin{equation}
\label{e41}
\hspace{1.0 cm} S(t) \sim \left( \frac{\kappa \; (1+\delta)}{\beta} \right)^{\frac{1}{1+ \delta}} \; e^{\left(\frac{\beta}{1+\delta}\right) \; t},
\end{equation}
in this case the expansion depends on $B$ parameter of the EoS only.
Thus at late universe we get the following
\begin{equation}
\label{e42}
\hspace{1.0 cm} \rho_{fb}+P_{fb} =0
\end{equation}
as given by eq.  (\ref{e40}) and from eqs. (\ref{e38}) and (\ref{e39}) we get
\begin{equation}
\label{e42a}
\rho +P_{r}=- \frac{(D-2)(D-3)}{2 r^{D-1}S^2} \; r_0^{D-3},
\end{equation}
\begin{equation}
\label{e42b}
\rho +P_{t}=\frac{(D-3)}{2 r^{D-1}S^2} \; r_0^{D-1},
\end{equation}
leading to $\rho+P_r < 0$ .  Consequently  although NEC is obeyed as the universe evolves out from wormhole throat, once again at the present epoch NEC and SEC do not  obey which accommodates the late time acceleration in the universe.


\begin{figure}
\includegraphics[scale=0.8]{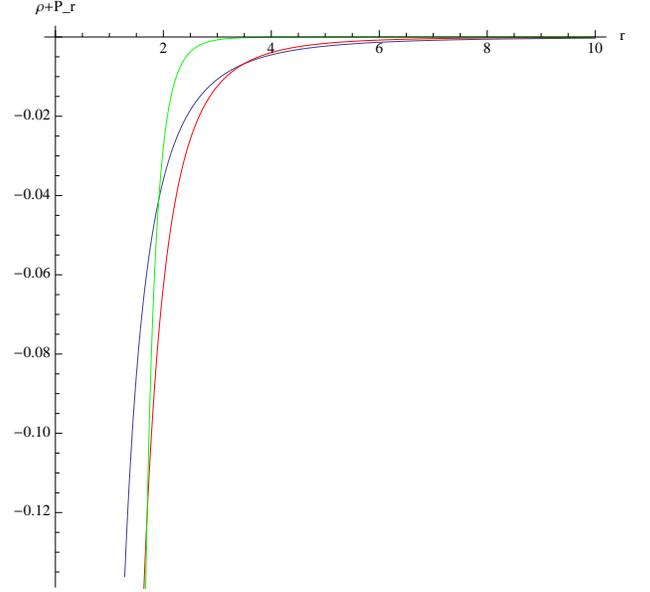}
\caption{Variation  of NEC ($\rho+p_{r})$ with  distance from the throat with $A=2$ and $B=1$ in different dimensions. The blue, red and green curves correspond to $D=4,5,10$ respectively.  }
\label{Fig: 5}
\end{figure}

\section{Higher Dimensional Flat Emergent Universe}

The conservation equation in a higher dimensional universe is given by
\begin{equation}
\label{e43}
\frac{d \rho }{dt} + (D-1) (\rho +P) \; \frac{\dot{S}}{S} =0.
\end{equation}
Integrating above eq. (\ref{e43}) using the non-linear equation  of state : $P=A \rho - B \sqrt{\rho}$, we get
\begin{equation}
\label{e44}
\rho(S)= \left( \frac{1}{1+A} \right)^2  \left( B+ \frac{K}{S^{\frac{(D-1)(1+A)}{2}}}\right)^2 
\end{equation}
where $K$ is an integration constant. The scale factor is a monotonically increasing function of cosmic time, thus the evolution of the universe can be studied making use of the energy density. Here the energy density depends on the equation of state parameters $A$, $B$, $D$ and an arbitrary integration  constant $K$. It is important to note here that the universe is composed of three different types of fluids. Thus the energy density given by eq. (\ref{e44}) can be expanded and we get
\[
\rho = \left(\frac{B}{1+A} \right)^2   + \frac{2\;B K}{(1+A)^2} \frac{1}{S^{\frac{(D-1) (1+A)}{2}}}
\]
\begin{equation}
\label{e45}
\hspace{2.5 cm} + \left( \frac{K}{1+A} \right)^2 \frac{1}{S^{(D-1)(1+A)}}.
\end{equation}
Using the energy density ($\rho$) in eq. (\ref{e35}), one gets the pressure which is given by
\[
P =-  \left(\frac{B}{1+A} \right)^2   + \frac{B K (A-1)}{(1+A)^2} \frac{1}{S^{\frac{(D-1) (1+A)}{2}}}
\]
\begin{equation}
\label{e46}
\hspace{2.5 cm} + \left( \frac{K}{1+A} \right)^2 \frac{1}{S^{(D-1)(1+A)}}.
\end{equation}
Thus the three types of fluids are expressed as
\begin{equation}
\rho=\rho_1+\rho_2+\rho_3
\end{equation}
\begin{equation}
P=P_1+ P_2+ P_3.
\end{equation}
Using the form of the barotropic equation : $P_i= \omega_i \rho_i$  with $i=1,2,3$ we obtain
\begin{equation}
\omega_1 =-1, \; \; \; \omega_2=\frac{A-1}{2}, \;\;\; \omega_3=A,
\end{equation}
One may now try to identify the components from the EOS of individual components:
The first term behaves like a cosmological constant $\Lambda = \frac{B}{(A+1)^2}$ which accounts for the dark
energy. The pressure of other fluids evolve as $P_2=  \frac{\omega_2}{ S^{\frac{(D-1)(A+1)}{2}}  }$  in unit of $\Lambda \left(\frac{K}{B} \right) $ and $ P_3=   \frac{\omega_3}{ S^{(D-1)(A+1)}} $ in unit of $\Lambda   \left( \frac{K}{B}  \right)^2$ which are determined by the  EoS parameter ($A$) and  dimensions ($D$). This is interesting to note that three different cosmic fluids are determined by a single  EoS $P= A \; \rho - B \; \sqrt{\rho}$ which is non-linear as given by  eq. (\ref{e35}).

Thus a higher dimensional emergent universe scenario can be constructed with  cosmic fluids which are presented in Table-1.
    
\begin{table}
 \centering
  \begin{tabular}{|c|c|c|c|c|c|}\hline
$A$ & $\frac{\rho_2}{\Lambda}$ in $\frac{K}{B}  $ & $\omega_2$ & $\frac{\rho_3}{\Lambda}$ in $\left(\frac{K}{B}\right)^2$ & $\omega_3 $ & Composition of fluids  \\ \hline
$\frac{1}{3}$ & $\frac{9}{8 }\; S^{-\frac{(D-1)}{3}}$ & $-\frac{1}{3 }$ & $\frac{9}{8 } \; S^{-\frac{4}{3} (D-1)}$ & $\frac{1}{3}$ & DE, EM, Radiation \\ \hline
$- \frac{1}{3} $ & $\frac{9}{2} \;  S^{- \frac{(D-1)}{3}}$ & $-\frac{2}{3 }$ & $\frac{9}{4} \;  S^{\frac{-2(D-1)}{3}}$ & $-\frac{1}{3}$ & DE, EM, Cosmic strings \\ \hline
$1 $ & $\frac{1}{2} \; S^{-(D-1)}$ & $0$ & $\frac{1}{4} \;  S^{-2 (D-1)}$ & $1$ & DM, EM, Stiff matter \\ \hline
$0 $ & $\frac{2}{8}\;  S^{- \frac{(D-1)}{2}}$ & $-\frac{1}{2 }$ & $S^{-(D-1)}$& $0$ & DE, EM,  Dust \\ \hline
\end{tabular}
\caption{Composition of cosmic matter for a choice of  $A$ in $D$ dimensions. The value of $\omega_2$ defines the exotic
matter (EM) with density $\rho_2$, Dark matter (DM) in addition to barotropic fluid  relevant in each case.}
 \label{tab:1}
\end{table} 

It is noted that the behaviour of the cosmic fluids in $D=3, \; 5, \; 7$ with $A=1, \; 0, \;  -\; \frac{1}{3}$ respectively matches with the time evolution that one gets in$D=4$ dimensions.The  cosmic evolution in  case $A= \frac{1}{3}$ is unique  only in $D=4$ dimensions. 

One of the problem is that if $A$ is specified for a known  dimensions of the universe $D$, the composition of the cosmic fluids are fixed in the case of non-interacting fluids. A physically acceptable emergent universe results if one considers interaction among the fluids. In the next section we consider interacting fluids.
 
\section{Interacting Cosmic Fluids}

We consider an emergent universe with non-linear EoS given by eq. (\ref{e35}), where an interaction among the fluid components.  In the conservation eq. (\ref{e43}), assuming an onset of interaction among the
cosmic the fluids at $t \geq t_o$,  the conservation equations for the energy densities of the
fluids now can be written as 
\begin{equation}
\label{e47a}
\frac{d \rho_1 }{dt} + (D-1)  H \;  (\rho_1 + P_1) = - Q', 
\end{equation}
\begin{equation}
\label{e47b}
\frac{d \rho_2 }{dt} + (D-1) H  \;  (\rho_2 + P_2) = Q',
\end{equation}
\begin{equation}
\label{e47c}
\frac{d \rho_3 }{dt} + (D-1) H \;  (\rho_3 + P_3) =  Q' - Q,
\end{equation}
 where  $H=\frac{\dot{S}}{S}$,  $Q$ and $Q' $ represent the interaction terms, which can have arbitrary form,  $\rho_1$  represents
the dark energy density, $\rho_2$ represents exotic matter, and $\rho_3$ represents normal matter. 
Here  $Q  < 0 $ corresponds to energy transfer from the exotic matter sector to two other
constituents, $Q' > 0 $ corresponds to energy transfer from the dark energy sector to the other
two fluids, and $Q'  < Q $ corresponds to energy loss for the normal matter sector. In the above 
 the limiting case $Q  = Q' $ represents that  dark energy interacts only with the exotic
matter. It is evident that although the above three equations are different, the total energy of
the fluid satisfies the conservation equation together which is given by eq. (\ref{e35}). It is possible to construct equivalent effective uncoupled equations described by the following conservation equations:
\begin{equation}
\label{e48a}
\frac{d \rho_1 }{dt} + (D-1) H \;  \left(1+ \omega_1^{eff} \right) \rho_1,
\end{equation}
\begin{equation}
\label{e48b}
\frac{d \rho_2 }{dt} + (D-1) H \;  \left(1+ \omega_2^{eff} \right) \rho_2,
\end{equation}
\begin{equation}
\label{e48c}
\frac{d \rho_3 }{dt} + (D-1) H  \;  \left(1+ \omega_3^{eff} \right) \rho_3,
\end{equation}
where the effective EoS parameters are given below:
\begin{equation}
\label{e49}
\omega_1^{eff} = \omega_1 + \frac{Q'}{(D-1) H \rho_1},
\end{equation}
\begin{equation}
\label{e49a}
\omega_2^{eff} = \omega_2 - \frac{Q}{(D-1) H \rho_2},
\end{equation}
\begin{equation}
\label{e49b}
\omega_3^{eff} = \omega_3 + \frac{Q-Q'}{(D-1) H \rho_3}
\end{equation}
Representing  the interaction as  $Q - Q' = - \; \beta H \rho_3,$ where $\beta$ is a coupling constant.  The effective state parameter for the normal fluid now can be expressed as
\begin{equation}
\label{e49c}
\omega_3^{eff} = \omega_3 -  \frac{\beta}{(D-1)}
\end{equation}
In figure 1, we plot the variation of effective EoS parameter  $\omega_3^{eff} $ with $ \omega_3$  (which corresponds to the EoS parameter $A$) for different strengths of interaction determined by $\beta$.  In Figs. (1) - (2) we plot $\omega_3^{eff} $ $vrs. $  $ \omega_3=A $ for different values of $D$ and $\beta$ respectively. It is evident from  Fig. (1) that for a given interaction the universe attains matter dominated phase  for which $\omega_3^{eff}=0$ faster than  the usual four dimensions. In Fig. (2), it is evident that for a given dimension of the universe, say $D=10$, the universe attains matter dominated universe faster for fluid with less interaction.
\begin{figure}[t]
\includegraphics[scale=0.8]{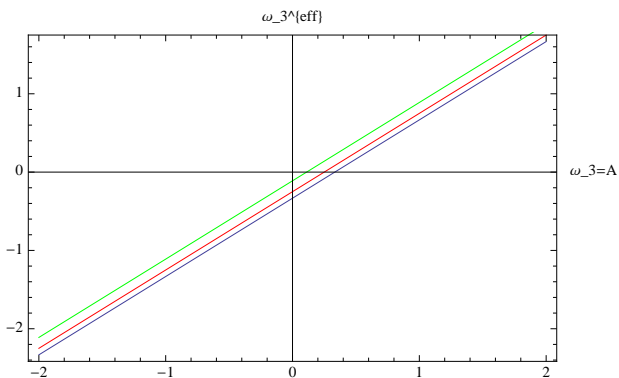}
\caption{Variation of $\omega_3^{eff} $ with $ \omega_3$ for $D=4$ (Blue), $D=5$ (Red), $D=10$  (Green) with $\beta=1$.}
\end{figure}
\begin{figure}[t]
\includegraphics[scale=0.8]{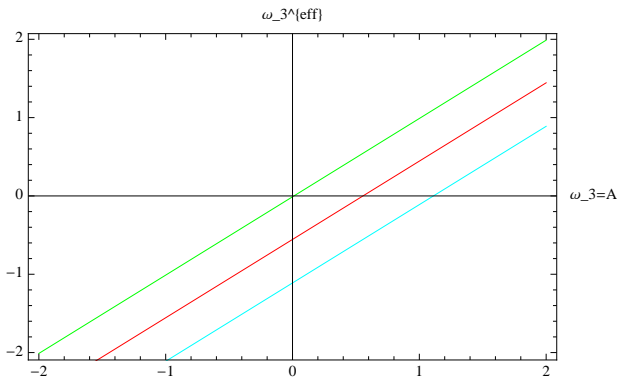}
\caption{Variation of $\omega_3^{eff} $ with $ \omega_3$ for  $\beta=10 $ (Cyan), $ \beta=5$ (Red),  $\beta=0.1$ (Green)  for $D=10$.}
\end{figure}

It is noted that as the strength of interaction ($\beta$) is increased, the value of  $\omega_3$  (i.e., $A$) for which $\omega_{eff} =0$ (corresponds to matter-domination) is found to increase. Thus a universe with any Eos parameter $A$ value is found to admit a matter-dominated phase at a late epoch, depending on the strength of the
interaction that was not permitted in the absence of interaction in the EU model as obtained earlier \cite{eu4}. Note that in the very early epoch, a universe is assumed to have a composition of three different fluids with no interaction in this picture at a later epoch it emerge out to a matter dominated phase with all the observed features at the present time. It is important to mention here that $A$ and $B$ determine the initial value of the cosmological constant $\Lambda$, but  at a later epoch when the interaction sets in $\Omega_{eff} \rightarrow \; -\; 1$ with the change in values of the interaction incorporated in  $Q' $ due to energy transfer  from one sector of matter to another and we also note that : $A_{eff} = A- \frac{\beta}{D-1} $. The interaction term $\beta$ is playing an important role.
\\

{\bf B. Higher Dimensional Closed or Open  Universe}\\

For a large radial distance $r$ away from the throat of the wormhole, one obtains  in Case (I)  $\alpha(r) \rightarrow k r^2$ which leads to open universe for $k=-1$ and in Case (II) $\alpha(r) \rightarrow k r^2$ which leads to a closed universe for $k=1$. 
The later solution is new which  permits two closed universe connected by a wormhole. Considering  a closed universe $i.e.$, $k=1$ we study the NEC condition to obtain dynamical wormhole which allows an emergent universe scenario as discussed above.
The shape function obtained from eq. (\ref{e17}) is given by 
\begin{equation}
\hspace{1.9 cm} b(r)= \frac{r_0^{D-3} + r_0^{D-1}}{r^{D-2}} - r^3.
\end{equation}


In Fig. (7) a 3D plot of radial pressure is plotted with radial distance and time and the variation of NEC is plotted  in Fig. (8) in $D=4$ dimensions, it is found that although the radial pressure is negative near the throat of the wormhole the NEC is obeyed. It is found that for asymptotic  closed universe, exotic matter is not required.

In Case  (II) which admits asymptotic open universe, $i.e.$ $k=-1$, the shape function obtained  from eq. (\ref{e17}) is given by 
\begin{equation}
\hspace{1.9 cm} b(r)= \frac{r_0^{D-3} + r_0^{D-1}}{r^{D-2}} - r^3.
\end{equation}
It has the same structure as obtained in closed universe which can be analyzed and in this case also one obtains wormholes without exotic matter.

\section{Constraints on the Tension and mass density at the throat}

The constraints on the tension and the mass density can be obtained from the higher dimensional Einstein field equation. At the wormhole throat $b=b_o$ and absence of a horizon, the field equation given by eq. (\ref{e12}) can be used to determine the tension in the throat of the wormhole which is given by
\begin{equation}
\label{e50}
\tau = - \; P_r (r, t)=\frac{(D-2)(D-3)}{16 \pi G_D c^4 \;  b_o^2 S_o^2(t)},
\end{equation}
where  the gravitational coupling constant $G_D= \frac{1}{M_{D}^{D-2}}$. It  is evident that the tension depends on the space-time dimensions and the size of the universe in the Einstein static phase. It may be important to mention here that in the emergent universe the initial size of the universe may be bigger than that one gets in the Planck regime of the FRW-universe. Following the  Brane world scenario and higher dimensions \cite{rm,ca}, it is found that
the $M_P \; G_4^{-\frac{1}{2}} \sim 1.22 \times 10^{16} \; TeV$, the tension at the throat of the wormhole from which emergent universe that  emerged is given by 
\begin{equation}
\label{e51}
\tau = \frac{(D-2)(D-3)\; c^4 }{16 \pi G \;  b_o^2 S_o^2},
\end{equation}
where $S_o=\left(\frac{(D-1)\;A \; k_o}{2} \right)^{\frac{2}{(D-1)\;A}}.$ Thus the equation of state parameter $A$, the arbitrary integration constant $k_o$ and dimensions of the universe $D$ are playing an important role.
It is evident that for a large size or the scale factor of the universe ($S_o$), the tension is very small. In four dimensions one gets tension at the throat which is given by
\begin{equation}
\label{e51a}
\tau = \frac{(D-2)(D-3) }{16 \pi G \; c^{-4} \;  b_o^2 \; S_o^2} \sim 5\times 10^{41} \frac{dyne}{cm^2} \left( \frac{10 \; m}{b_o \; S_o} \right)^2.
\end{equation}
The tension depends  on the initial size of the universe in the case of an Emergent Universe. In emergent universe scenario the size of a Einstein static universe $S_o$ is very big and $D$-dimensional Newtonian constant ($G$) is large  which reduces the tensions considerably, thus it is an important property of a Higher dimensional universe to  play an important role for understanding migration from one universe to the other dual space in the case  $\frac{\partial R}{\partial r}=0$. For $D=3$, the effective tension at the throat vanishes, thus it gives rise to an effective phenomena where the tidal force is absent at the throat. Thus an interstellar migration is possible without any hindrance if the space-time dimension is lower than the usual four dimensions. 
The mass function is $ M =  \rho V_D $
where the volume  is  $ V_D= \frac{ 2 \pi^{\frac{1}{2} (D-1) } }{ (D -1) \Gamma ((D -1)/2)}\; r^{D-1}$ . 
It is also evident that wormhole with two open universes connected at the throat can be obtained where NEC is  obeyed always.

\section{Discussion}

The wormhole solutions which permits emergent universe models are obtained in Einstein gravity in higher dimensions. In this framework it is shown that the initial Einstein static universe at infinite past in an emergent universe scenario can be represented by a dynamical wormhole throat. The dynamical evolution of the wormhole encompasses the flat emergent universe with  all the observed features. Considering a particular class of wormhole solutions corresponding to a spatially homogeneous Ricci scalar we determine the shape functions  for the wormhole in a flat, closed and open universe.   The cosmological solutions  in higher dimensions accommodating emergent universe is obtained here for homogeneous matter with non-linear  EoS.  We note that EU model  can be obtained for flat, asymptotically closed or open universe in  higher dimensions. For $D=4$, the EU model obtained by us \cite{eu4} is recovered. It is shown that  a dynamical wormhole joins the throat of the wormhole with the present universe. The initial Einstein's static universe required for EU scenario is the throat of the wormhole.  
 It is shown in Fig (1) that the size of the  initial static Einstein universe is determined by the EoS parameter $A$. 
 It is also noted that the maximum size  of the Einstein static universe  does not  depend on   the spacetime dimensions and geometry of the universe. 
 In Figs. (3) and (4), the variation of NEC is studied in a flat universe with different  values of the  parameters $A$ and $B$ in $D$ dimensions. It is found that  at the throat NEC is violated but away from the throat  it is obeyed.  In Fig. (5) it is evident that in a four dimensional universe although NEC violates near the throat but it begins to obey NEC near to the throat compared to a universe with dimensions more than four dimensions.  In Figs. (6) and (7), we note that the observed  universe is obtained in the case of interacting fluids which is determined by the  coupling term $\beta$ of the interaction although initially the composition of the universe is determined by $A$.
Another interesting solution is that for a closed or open universe dynamical wormholes are permitted with matter that respect the NEC in four and in higher dimensions. It is a new result as earlier cosmological models  different from EU models are obtained in 5-dimensions that respect the energy conditions \cite{22a}.
 A new class of  wormhole solutions for closed model of the universe is obtained here for dynamical wormhole that leads to flat emergent universe.
 A  flat universe requires exotic matter to begin with near the throat but away from the throat no exotic matter is required for flat emergent universe scenario.  
 A detail numerical analysis is displayed in  Table-1 and Table-2.  It is evident from the  3D plots in Figs. (8) and (9),  that even if the radial pressure is negative in an asymptotically closed universe, NEC is obeyed implying wormhole solution without exotic matter. The tension at the throat of the wormhole is estimated in eq. (73), it is noted that the tension in a 4-dimensional universe depends on the throat of the wormhole ($S_o$). In EU model the size of the static Einstein universe may be large, thus in that case the tension reduces considerably.

\vspace{1.0 cm}

{\bf Acknowledgment}
The author would like to thank IUCAA . Pune for supporting a visit and IUCAA Centre for Astronomy Research and Development (ICARD), NBU for extending research facilities. BCP would like to thank DST-SERB Govt. of India (File No.:EMR/2016/005734) for a project.\\

 
\end{document}